\documentclass[submission]{eptcs}
\providecommand{\event}{LSFA 2023} 

\usepackage{iftex}

\ifpdf
  \usepackage{underscore}         \usepackage[T1]{fontenc}        \else
  \usepackage{breakurl}           \fi

\usepackage[T1]{fontenc}
\usepackage[dvipsnames]{xcolor}
\usepackage{fancybox}
\usepackage{graphicx}
\usepackage{amsmath,amssymb,bm}
\usepackage{booktabs}
\usepackage{ebproof}
\usepackage{hyperref}
\usepackage{enumitem}
\newtheorem{example}{Example}[section]
\usepackage{color}

\newtheorem{innercustomthm}{Lemma}
\newenvironment{customthm}[1]
  {\renewcommand\theinnercustomthm{#1}\innercustomthm}
  {\endinnercustomthm}
\newcommand{\Env}{\mathcal{E}}
  \newcommand{\B}{\mathbb{B}}
  \newcommand{\R}{\mathbb{R}}
   \newcommand{\V}{\mathbb{V}}
  \newcommand{\boolExprs}{\mathcal{B}}
  \newcommand{\realExprs}{\mathcal{R}}
  \newcommand{\ra}{\rightarrow}
 \newcommand{\dfn}{\triangleq}

\newcommand{\name}{Plaidypvs}
\newcommand{\asub}{\csf{assign\_sub}}
\newcommand{\csf}[1]{\textbf{#1}}

\newcommand{\pathc}{\csf{path}}

\newcommand{\nqB}{nqB}
\newcommand{\circc}{\csf{circ}}

\newcommand{\NQBr}{\csf{NQB\_rel}}
\newcommand{\rel}{rel}

\newcommand{\fresh}{\csf{fresh?}}

\newcommand{\with}{\csf{with}}

\newcommand{\TRUE}{\csf{True}}
\newcommand{\FALSE}{\csf{False}}

\newcommand{\sol}{\csf{sol?}}

\newcommand{\val}{\csf{val}}
\newcommand{\cnst}{\csf{cnst}}

\newcommand{\True}{\csf{True}}

\newcommand{\sr}{\csf{s\_rel}}
\newcommand{\srd}{\csf{s\_rel\_diff}}
\newcommand{\eat}{\csf{e\_at\_t}}

\newcommand{\SUB}{\csf{SUB}}
\newcommand{\SUBre}{\csf{SUB\_re}}

\newcommand{\dls}[1]{\hypertarget{#1}{{
\textcolor{Blue}{\textbf{#1}}}}}
\newcommand{\dll}[1]{\hypertarget{#1}{\textcolor{Blue}{\textbf{#1}}}}

\newcommand{\dlref}[1]{\textcolor{Blue}{\hyperlink{#1}{{\textbf{#1}}}}}

\newcommand{\DDL}{\text{\upshape\textsf{d{\kern-0.05em}L}}}

\makeatletter
\gdef\tshortstack{\@ifnextchar[\@tshortstack{\@tshortstack[c]}}
\gdef\@tshortstack[#1]{\leavevmode
  \vtop\bgroup
    \baselineskip-\p@\lineskip 3\p@
    \let\mb@l\hss\let\mb@r\hss
    \expandafter\let\csname mb@#1\endcsname\relax
    \let\\\@stackcr
    \@ishortstack}
\makeatother
\begin{document}
\title{Embedding Differential Dynamic Logic in PVS}

\author{J. Tanner Slagel  \institute{NASA Langley Research Center \\ Hampton, VA, 23666, USA} \email{j.tanner.slagel@nasa.gov} \and Mariano Moscato \institute{National Institute of Aerospace$^{\star}$ \\ Hampton, VA, 23666, USA}
 \and Lauren White \institute{NASA Langley Research Center \\ Hampton, VA, 23666, USA}
  \and C\'{e}sar A. Mu\~{n}oz \institute{NASA Langley Research Center\\ Hampton, VA, 23666, USA}
   \and Swee Balachandran \institute{National Institute of Aerospace\thanks{Institute at time of contribution.} \\ Hampton, VA, 23666, USA}
    \and Aaron Dutle \institute{NASA Langley Research Center \\ Hampton, VA, 23666, USA}} \def\authorrunning{JT Slagel et al.}
\def\titlerunning{Embedding Differential Dynamic Logic in PVS}
\maketitle              \begin{abstract}
Differential dynamic logic  (\DDL{}) is a formal framework for specifying and reasoning about hybrid systems, i.e., dynamical systems that exhibit both continuous and discrete behaviors. These kinds of systems arise in many safety- and mission-critical applications. 
This paper presents a formalization of \DDL{} in the Prototype Verification System (PVS) that includes the semantics of hybrid programs and {\DDL}'s proof calculus. The formalization embeds {\DDL} into the PVS logic, resulting in a version of \DDL{} whose proof calculus is not only formally verified, but is also available for the verification of hybrid programs within PVS itself. This embedding, called {\name} (\textbf{P}roper\textbf{l}y \textbf{A}ssured \textbf{I}mplementation of \textbf{d}L for
H\textbf{y}brid \textbf{P}rogram \textbf{V}erification and \textbf{S}pecification),
supports standard \DDL{} style proofs, but further leverages the capabilities of PVS to allow reasoning about entire classes of hybrid programs.
The embedding also allows the user to import the well-established definitions and mathematical theories available in PVS. 
 
\end{abstract}

\section{Introduction}

Systems that exhibit both discrete and continuous dynamics, known as \emph{hybrid systems}, have emerged in numerous safety- and mission-critical applications such as avionics systems, robotics, medical devices, railway operations, and autonomous vehicles. To formally reason about these systems, 
it is often useful to model them as \emph{hybrid programs} (HPs), where the discrete variables evolve through assignments like traditional imperative programs and the continuous variables are defined by a system of differential equations. Hybrid programs are suitable to model complex dynamics where the continuous and discrete dynamics are largely intertwined, but due to their complexity, efficient and effective formal reasoning about properties of such programs can be a challenge.

Differential dynamic logic (\DDL) enables the specification and reasoning of HPs using a small set of proof rules~\cite{DBLP:conf/tableaux/Platzer07,platzer2008differential,platzer2017complete,Platzer18}. Conceptually \DDL{} can be split into two parts: (1)~a framework for the logical specifications of HPs and their properties and (2)~a proof calculus that is a collection of axioms and deductive rules for reasoning about these logical specifications. The KeYmaera~X\footnote{\href{https://keymaerax.org}{https://keymaerax.org}}  theorem prover is a software implementation of \DDL{} built up from a small, trusted core that assumes the axioms of \DDL{}~\cite{fulton2015keymaera,mitsch2020retrospective,mitsch2021implicit} with a web-based interface for specification and reasoning of HPs~\cite{mitsch2017keymaera}. KeYmaera X has been used in the formal verification of several cyber-physical systems~\cite{DBLP:journals/sttt/JeanninGKSGMP17,DBLP:journals/ijrr/MitschGVP17,KabraMP22,DBLP:journals/tecs/CleavelandMP23,DBLP:journals/ral/BohrerTMSP19,DBLP:conf/aaai/FultonP18,DBLP:conf/rssrail/MitschGBGP17,DBLP:conf/fase/MullerMRSP17}. 

This paper presents a formal embedding of \DDL{} in the Prototype Verification System (PVS). PVS is a proof assistant that integrates a fully typed functional specification language supporting predicate subtypes and dependent types with an interactive theorem prover based on higher order logic. PVS allows users to write formal specifications and reason about them using a collection of built-in proof rules and user-defined proof strategies. Strategies are built on top of proof rules in a conservative way so that they do not introduce additional soundness concerns. Formal PVS developments are structured in theories and a collection of theories form a library. The NASA PVS Library (NASALib)\footnote{\href{https://github.com/nasa/pvslib}{https://github.com/nasa/pvslib}} is a collection of formal developments contributed by the PVS community and maintained by the Formal Methods Team at NASA Langley Research Center. Currently, it consists of over $38,000$ proven lemmas spanning across $69$ folders related to a wide range of topics in mathematics, logic, and computer science. The work presented in this paper relies on and contributes to NASALib.

The primary contribution of this work is a formal development called {\em \name{}} (\textbf{P}roper\textbf{l}y \textbf{A}ssured \textbf{I}mplementation of \textbf{D}ifferential Dynamic Logic for H\textbf{y}brid \textbf{P}rogram \textbf{V}erification and \textbf{S}pecification), which is publicly available as part of NASALib\footnote{\href{https://github.com/nasa/pvslib/tree/master/dL}{https://github.com/nasa/pvslib/tree/master/dL}}. {\name} includes the
{\em specification} of \DDL{}'s HPs and their properties through an embedding in the PVS specification language, the {\em verification} of correctness of {\DDL}'s axioms and deductive rules, and the {\em implementation} of these rules through the strategy language of PVS, resulting in a formally verified and interactive implementation of the proof calculus of \DDL{} within PVS. 

While reasoning about HPs using a formally verified implementation of \DDL{} is already an achievement, the integration in PVS brings additional opportunities for extending the functionality of \DDL{} beyond what is available in a stand-alone \DDL{} system such as KeYmaera~X. For example, new or existing functions and definitions in PVS can be used inside of the \DDL{} framework. This includes trigonometric and other transcendental functions already specified in NASAlib, as well as the corresponding properties concerning their derivatives and integrals. In addition, meta-reasoning about HPs and their properties can be performed in PVS using the \DDL{} embedding. Examples include specifying HPs with a parametric number of variables, which can be used to reason about situations with an unknown but finite number of actors, and reasoning about entire classes of HPs, which can be specified using the PVS type system.

The rest of this paper proceeds as follows. Section~\ref{sec:HP} details the formal development of HP specifications in \name. Section ~\ref{sec:DDL} gives an overview of the formal verification approach to prove \DDL{} statements in PVS, as well the implementation of the proof calculus of \DDL{} in the PVS prover interface. Section~\ref{sec:using} shows an example of utilizing the features of \name{} beyond the capabilities of \DDL{} alone. Related work is discussed in~\ref{sec:related}. Finally, conclusions and future work are discussed in~\ref{sec:con}.

  \section{Specification of hybrid programs} \label{sec:HP}
 This section describes the syntax, semantics, and logical specifications of HPs developed in \name{}. Before these are introduced, a few preliminary concepts are needed. \subsection{Environment, real expressions, Boolean expressions} \label{sec:env}
Hybrid programs manipulate real number values using discrete and continuous operations. At any moment, the state of a hybrid program is given 
by an environment of type $\mathcal{E} \dfn [\mathbb{V} \rightarrow \mathbb{R}]$ that maps program variables in $\mathbb{V}$ to real number values in $\mathbb{R}$, where $\mathbb{V}$ is an infinite, but enumerable set of variables and $\mathbb{R}$ is the set of real numbers. For simplicity, variables are represented by indices, i.e., $\mathbb{V}$ is just the set of natural numbers.

The sets $\mathcal{R}$ and $\mathcal{B}$ of real and Boolean hybrid program expressions, respectively, are defined by a shallow embedding meaning they are represented by their evaluations functions, i.e.,
$\mathcal{R} \dfn  [ \mathcal{E} \rightarrow \mathbb{R}]$ and 
 $\mathcal{B} \dfn    [\mathcal{E} \rightarrow \mathbb{B}].$ For instance, $\cnst(c) \dfn \lambda(e:\mathcal{E}). c$ represents the constant expression that returns the value $c \in \mathbb{R}$ in any environment and $\val(v) \dfn \lambda(e:\mathcal{E}). e(v)$ represents the real expression that returns the value of variable $v$ in the environment $e$.
Similarly, $\top \dfn \lambda(e:\Env). \TRUE$ and $\bot \dfn \lambda(e:\Env). \FALSE$ represent the Boolean hybrid program constants that always return $\TRUE \in \B$
and $\FALSE \in \B$, respectively.
While real and Boolean expressions can be arbitrary functions, \name{} provides support for standard arithmetic and Boolean operators by lifting them to the domain of $\realExprs$ and $\boolExprs$.
Given $r, r_{1},r_{2} \in \mathcal{R}$ and $n \in \mathbb{N}$ the following are recognized to be of type $\realExprs$: $r_{1} + r_{2}$, $r_{1} - r_{2}$, $r_{1}/r_{2}$, $r_{1}\cdot r_{2}$, $r_1 = r_2$, $-r$, $\sqrt{r}$, and $r^{n}$.
It is important to notice that, for instance, in the real expression $r_1 + r_2$, the operator $+$ is not the arithmetic addition, but it is of type $\realExprs \times \realExprs \ra \realExprs$.
Similarly, given Boolean expressions $b, b_{1},b_{2} \in \mathcal{B}$, the following are recognized to be of type $\boolExprs$: $b_{1} \wedge b_{2}$, $b_{1} \vee b_{2}$, $b_{1} \rightarrow b_{2}$, $b_{1} \leftrightarrow b_{2}$, and $\neg b$.

\begin{example}[Environments, Real and Boolean Expressions] \label{ex:reex}
Let $x,y \in \mathbb{V}$ and $c  \in \mathbb{R}_{\geq 0}$, the following Boolean expression denotes a circle of radius $c$ centered at $(0,0)$:
\begin{equation}
    \label{eq:circ}
    \val(x)^2 + \val(y)^2 = \cnst(c)^2.
    \end{equation}
Furthermore, assuming the environment $ e  \dfn (\lambda(v\, : \, \mathbb{V}). 0) \,\, \with{}\,\, \{x \mapsto c/2, y \mapsto \sqrt{3}\cdot c / 2\}$, the following Boolean statement holds.
$$(\val(x)^2 + \val(y)^2 = \cnst(c)^2)(e) =  \TRUE.$$
\end{example}
Henceforth, for ease of presentation, the $\val$ and $\cnst$ operators are suppressed in much of the remainder of the paper. The Boolean expression in Formula~\ref{eq:circ}, for example, will be presented instead as $x^2+y^2 = c^2.$

\subsection{Hybrid programs}
Hybrid programs are syntactically defined as a datatype $\mathcal{H}$ in PVS according to the following grammar.
\begin{align*}  \alpha ::=\ & \mathbf{x} := \ell \  \rvert\ \mathbf{x}' = \ell\, \& \,P\ \rvert\  ?P\ \rvert\  x := * \ \rvert\  \alpha_1 ;\alpha_2\ \rvert\  \alpha_1 \cup \alpha_2\ \rvert\  \alpha_{1}^{*}. \end{align*}
 Here, $\mathbf{x} := \ell$ is a list of pairs in $\mathbb{V}\times \mathcal{R}$, where the first entries are unique, intended to represent a discrete assignment of the variables indexed by these first elements. The differential equation $\mathbf{x}' = \ell\, \&\,P$,
where $\mathbf{x}' = \ell$ is another such list in $\mathbb{V}\times \mathcal{R}$ and $P \in \mathcal{B}$ is a Boolean expression,
is meant to symbolize the continuous evolution of the variables in $\mathbf{x}'$ according to the first order differential equation described by $\ell$. Note that use of the symbol $\&$ is distinct from Boolean conjunction and is used here purely syntactically to represent that
the solution of the differential equation satisfies $P$ along the evolution. To reference a variable used in a discrete assignment or differential equation, the notation $v \in \mathbf{x}$ (respectively, $v \in \mathbf{x}'$) will be used. The real expression associated with $v$ in $\ell$ will be denoted $\ell(v)$. The program $?P$ represents a check of the Boolean expression $P$. The program $x := *$ represents a discrete assignment of the variable $x$ to an arbitrary real number value.
 The program $\alpha_1 ;\alpha_2$ represents the sequential execution of the sub-programs $\alpha_1$ and $\alpha_2$,
 while $ \alpha_1 \cup \alpha_2$ symbolizes a nondeterministic choice between two subprograms. Finally, $\alpha^{*}_{1}$ represents repetition of a HP a finite but unknown (possibly zero) number of times.

Formally, the predicate $\sr$ defines the semantic relation of a hybrid program $\alpha$ with respect to input and output environments $e_{i},e_{o} \in \mathcal{E}$. It is inductively defined on $\alpha$ as follows. 
\[\sr(\alpha)(e_{i})(e_{o}) \dfn \begin{cases}
\forall k:\, k\notin \mathbf{x} \rightarrow e_o(k)=e_i(k) & \textrm{if } \alpha = (\mathbf{x} := \ell),   \\
\hspace{19 pt} \wedge \, k \in\mathbf{x} \rightarrow e_o(k) = \ell(k)(e_i)  & \\
e_{o} = e_{i} \, \vee \,   & \textrm{if } \alpha = (\mathbf{x}' = \ell\, \& \,P), \\
 \hspace{18 pt}  \exists D : \, \srd(D,\mathbf{x}',\ell,P,e_{i},e_{o}) & \\
e_{o}=e_{i} \wedge P(e_{i}) & \textrm{if } \alpha = ?P, \\
\exists r :  e_{o}(x) = r \wedge Q(r)(e_{i})  & \textrm{if } \alpha = (x := * \, \, \& \, Q), \\
\exists e : \sr(\alpha_{1})(e_{i})(e)  & \textrm{if } \alpha = \alpha_1 ;\alpha_2,  \\
\hspace{18 pt} \wedge \, \sr(\alpha_{2})(e)(e_{o}) &  \\
\sr(\alpha_{1})(e_{i})(e_{o})  & \textrm{if } \alpha =  \alpha_1 \cup \alpha_2, \\
 \vee \,  \sr(\alpha_{2})(e_{i})(e_{o})  &  \\
 e_{o} = e_{i} \, \vee \,   & \textrm{if } \alpha =  \alpha_{1}^{*}.  \\
  \hspace{18 pt}  \exists e : \sr(\alpha_{1})(e_{i})(e)  & \\
  \hspace{18 pt} \hspace{18 pt}  \wedge \,  \sr(\alpha)(e)(e_{o}) &
\end{cases} \]
The correspondence between the informal description of semantics and the $\sr$ function is standard in all cases except the differential equation branch. For differential equations, the domain $D$ is $\mathbb{R}_{\geq0}$, or some closed interval starting at $0$, and the semantics is given by the following function.
\begin{align*}
\srd(D, \mathbf{x}',\ell,P,e_{i},e_{o}) \ \dfn \  &  \exists r : \exists! f : \, D(r) \wedge \sol( D,\mathbf{x}', \ell,e_{i})(f) \wedge  \\
& \quad  e_{o} = \eat(\mathbf{x}',\ell,f,e_{i})(r) \, \wedge \\
& \quad  \forall t: \, \left(D(t) \wedge t \leq r \right) \\ &  \quad  \quad \rightarrow P(\eat(\mathbf{x}', \ell,f,e_{i})(t)).
\end{align*}
Unpacking this further, $$\eat(\mathbf{x}', \ell,f,e_{i}) \ \dfn\ \lambda(r: \mathbb{R}).\lambda(j: \mathbb{V}). \begin{cases} e_{i}(j) & \textrm{if }  j \notin\mathbf{x}', \\ f(j)(r)
  & \textrm{if } j \in\mathbf{x}', \end{cases} $$ is a function that characterizes the environment $e_i$, with the continuously evolving variables $\mathbf{x}'$ replaced by values from a function $f : [ \mathbb{R}^k
 \rightarrow \left[\mathbb{R} \rightarrow \mathbb{R}\right]].$  The definition
\begin{align*} \sol(D,\mathbf{x}',\ell,e_{i})(f) \ \dfn\  & \forall(i \in \mathbf{x}',
                                                     t \in \mathbf{D}): \\
& (f(i))'(t) = \ell(i)(\eat(\mathbf{x}', \ell, f,e_{i})(t))  \end{align*}
ensures that $f$ is the solution to the $k$-dimensional differential equation $\mathbf{x}' = \ell$ throughout the domain $D$. Note in the definition of $\srd$ this solution $f$ is further assumed to be unique on the domain $D$.

Below is a colloquial description of the semantics of each type of HPs, where $e_{i},e_{o}$ are the input and output environments, respectively.
\begin{itemize}[align=parleft,leftmargin=*,labelindent=0mm,labelsep=20mm]
\item[$\mathbf{x}:=~\ell$] Discrete variable assignment. This means that $e_{i}$ and $e_{o}$ agree on all the variables not mentioned in $\ell$, and for the variables in $\ell$, a discrete jump has taken place.
\item[$\mathbf{x}':=~\ell\, \& \, P$] Continuous variable assignment. Continuous jumps take place where the output variable that is included in $\ell$ has evolved according to the first order differential equation defined in $\ell$. The solution to the differential equation satisfies $P$.
\item[$?P$] Test HP. An input/output pair is related only if they are equal and $e_{i}$ satisfies $P$.
\item[$\ x := ~*$] Random discrete assignment. Random assignment of variable $x$, where the random assignment is some value $r$ and $e_{o}(n) = r$.
\item[$\alpha_{1} ; ~\alpha_{2}$] Sequential HP. Runs two HPs $\alpha_{1}$ and $\alpha_{2}$ in order such that there is an environment $e$ that is semantically related to $e_{i}$ through $\alpha_{1}$, and semantically related to $e_{o}$ through $\alpha_{2}$.
\item[$\alpha_{1} \cup ~\alpha_{2}$] Nondeterministic choice HP. This HP nondeterministically chooses one of $\alpha_{1}$ or $\alpha_{2}$. Here $e_{o}$ is semantically related to $e_{i}$ through $\alpha_{1}$ or $\alpha_{2}$.
\item[$\alpha^{*}$] Loop HP. This is the repeat of HP $\alpha_{1}$ a finite but undisclosed number of times. The environment $e_{o}$ is either equal to $e_{i}$ or is it semantically related to another environment $e$ through $\alpha$ and $e$ is semantically related to $e_{o}$ through $\alpha^*$.
\end{itemize}

\begin{example}[HP] \label{ex:hpe}
The hybrid program
\begin{align*} 
  ((&?(x > 0); (x' = -y, y' = x\  \& \, x \geq 0)) \, \cup  \\
& (?(x \leq 0); (x' = -c, y' = 0)))^{*}, \end{align*}
where $x,y \in \V$, $x \neq y$, and $c \in \R$,
represents the dynamic systems where $x$ and $y$ progress according to the differential equation $x' = -y,$ $y' =x$ when $x > 0 $, but when $x \leq 0$ the variables progress according to the differential equation $x' = -c,$ $y'=0$. Note that the test statements, introduced by the operator~$?$, determine which branch of $\cup$ in the HP is applicable, and the domain $x \geq 0 $ in the first differential equation prevents the dynamics from continuing when $x=0$, forcing the other branch of the HP to take place. The operator~$*$ allows repetition so that both branches of the dynamics are carried out.
\end{example}

\begin{figure} 
\begin{center}
\includegraphics[width=.5\textwidth]{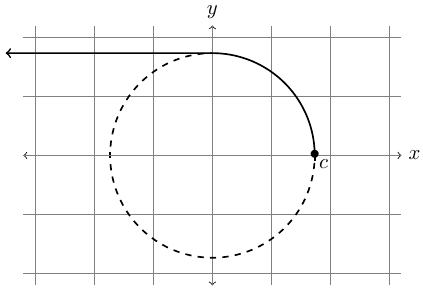}
\end{center}
\caption{Dubins path modeling an aircraft turning.}
\label{fig:dubf}
\end{figure}

The hybrid program in Example~\ref{ex:hpe} will be used as running example through this paper. It models a Dubins curve representing the trajectory of an aircraft turning and then proceeding in a straight line (see Figure~\ref{fig:dubf}).

\subsection{Quantified statements about hybrid programs}
A hybrid program can have potentially many different executions or \textit{runs}. This means that given an input environment $e_{i}$, there may be infinitely many output environments $e_{o}$ semantically related to it (by repetition, random assignment, etc.).
To reason about these runs, universal and existential quantifiers over the potentially infinite number of executions of an HP are defined. These quantifies are called {\em allruns}, denoted $\left[ \, \cdot \, \right]$, and {\em someruns}, denoted $\langle\, \cdot \, \rangle$. For $\alpha \in \mathcal{H}$ and $P \in \mathcal{B}$, $\left[\alpha\right]P \in \mathcal{B}$ is defined as follows.
$$\left[\alpha\right]P \ \dfn\ \lambda(e_{i}: \mathcal{E}). \forall e_{o} : \sr(\alpha)(e_{i})(e_{o}) \rightarrow P(e_{o}),$$
Analogously, $\left\langle \alpha \right\rangle P \in \mathcal{B}$ is defined as follows.
$$\left\langle \alpha \right\rangle P \ \dfn\ \lambda(e_{i}: \mathcal{E}). \exists e_{o} : \sr(\alpha)(e_{i})(e_{o}) \wedge P(e_{o}).$$
These quantifiers state that every (some, respectively) run of the HP $\alpha$ starting at environment $e_i$ and ending at environment $e_o$ satisfies~$P.$
\begin{example}[Allruns] \label{ex:ale}
Let $\alpha$ be the HP in Example \ref{ex:hpe}, $\circc(c) \dfn x^2 + y^2 =c^2$ and
\begin{align*} \pathc(c) \dfn (x > 0\ {\rightarrow}\ \circc(c)) {\ \wedge\ }
 (x \leq 0 {\ \rightarrow\ } y = c). \end{align*}
 Then, the Boolean expression 
 \begin{align}
\label{eqn:ale}
   (x = c {\ \wedge\ } y = 0) {\ \rightarrow\ } \left[\alpha\right] \pathc(c),
   \end{align}
states that if the value of $x$ is $c$ and the value of $y$ is $0$, then for all runs of the HP $\alpha$, the values of $x$ and $y$ stay inside $\pathc(c)$. In other words, $x$ and $y$ stay on the circle of radius $c$ until $x=0$ and then stay on the line $y=c$.
\end{example}

 \section{Embedding differential dynamic logic} \label{sec:DDL}
With the formal specification of hybrid programs established, the
embedding of the sequent calculus of \DDL{} in PVS can be
discussed. First,  \DDL{}-sequents will be defined, then a
description of the formal verification process encoding the axioms and
rules of \DDL{} as lemmas in PVS is provided.

\subsection{\DDL{}-sequents} \label{sec:seq}
A \DDL{}-sequent is denoted $\Gamma \vdash \Delta$, where $\Gamma$ and $\Delta$, known
as the {\em antecedent} and the {\em consequent}, respectively, are
lists of Boolean expressions. In PVS, a \DDL{}-sequent is defined by
\begin{align*}
\Gamma \vdash \Delta \ \dfn\ \forall e \in \mathcal{E} :   \bigwedge \Gamma(e)  \implies \bigvee \Delta(e),
\end{align*}
where $\implies$ is the PVS implication.
Intuitively, this means that the conjunction of the antecedent formulas
implies the disjunction of the consequent formulas.

The \DDL{} approach for proving statements about hybrid programs relies
on a set of deductive rules of the form
\[\begin{prooftree}
  \hypo{\Gamma_1 \vdash \Delta_1 \ \ \ \ \ldots \ \ \ \ \Gamma_k \vdash \Delta_k}
  \infer1
    {\Gamma \vdash \Delta.}
 \end{prooftree}\]
Rules without hypothesis, i.e., where $k=0$, are called {\em axioms}.
A rule of this form states that the conjunction of the sequents above
the inference line implies the sequent below the inference line. When
proving statements, these rules are used in a bottom-up fashion forming an
inverted (proof) tree, where the root of the tree is the sequent to be
proven, branches are related by instances of
deductive rules, and leaves are instances of axioms.

To formally verify \DDL{}
each
rule of \DDL{} is specified as a PVS lemma, which takes essentially the
following form.

\begin{customthm}{$<$\DDL{}-rule-name$>$}
  For all lists of Boolean expressions $\Gamma, \Delta$,  
  $$\bigwedge_{i=1}^k \Gamma_i \vdash \Delta_i\ \  \implies\ \ \Gamma
\vdash \Delta. $$
\end{customthm}

With such lemmas proven in PVS, a user can bring them into a
proof environment and instantiate them as needed for proving a
specific sequent. 
To automate this process, these lemmas are further implemented as
\emph{(proof) strategies} in
PVS. These strategies parse the current sequent, identify instantiations that apply, 
hide unneeded formulas, and prove type-checking conditions that may
appear, among other capabilities. More complex strategies are built on
top of these strategies to simplify the proof process. 
Some of these
rules, including details about their specification, verification, and
implementation as strategies in PVS, are discussed below. A \name{} ``cheat sheet'' is available for users with the development.\footnote{\href{https://github.com/nasa/pvslib/tree/master/dL/cheatsheet.pdf}{https://github.com/nasa/pvslib/tree/master/dL/cheatsheet.pdf}}

\subsection{Basic logical and structural rules of \DDL{}} \label{ref:log}
  \begin{figure}[t]
  
\noindent\ovalbox{
{ \small
\begin{minipage}[t]{0.45\textwidth}
\begin{tabular}{rl}
\dls{notR} & \begin{prooftree}
  \hypo{ \Gamma, P \vdash \Delta }
  \infer1
    { \Gamma \vdash \neg P, \Delta }
 \end{prooftree} \\[9 pt]
\dls{notL} & \begin{prooftree}
  \hypo{\Gamma \vdash P, \Delta }
  \infer1
    {\Gamma, \neg P \vdash  \Delta }
 \end{prooftree} \\[9 pt]
  \dls{andR} & \begin{prooftree}
  \hypo{\Gamma \vdash P, \Delta  \,\,\,  & \Gamma \vdash Q, \Delta}
  \infer1
    {\Gamma \vdash  P \wedge Q,  \Delta }
 \end{prooftree} \\[9 pt]
   \dls{andL} & \begin{prooftree}
  \hypo{\Gamma, P, Q \vdash \Delta}
  \infer1
    {\Gamma, P \wedge Q \vdash \Delta }
 \end{prooftree}  \\[9 pt]
    \dls{orR} & \begin{prooftree}
  \hypo{\Gamma \vdash P, \, Q, \, \Delta}
  \infer1
    {\Gamma \vdash P \vee Q, \, \Delta }
 \end{prooftree} \\[9 pt]
    \dls{orL} & \begin{prooftree}
  \hypo{\Gamma, \, P  \vdash \Delta \,\,\, & \Gamma, \, Q \vdash \Delta}
  \infer1
    {\Gamma,\, P \vee Q \vdash \Delta }
 \end{prooftree} \\[9 pt]
 \dls{cut} & \begin{prooftree}
  \hypo{\Gamma \vdash C, \, \Delta \, \, \,& \Gamma, \, C \vdash \Delta}
  \infer1
    {\Gamma \vdash \Delta }
    \end{prooftree}
    \\[9 pt]
   \dls{weakR} & \begin{prooftree}
  \hypo{\Gamma \vdash P, \Delta \, \, \,& P \vdash Q}
  \infer1
    {\Gamma \vdash Q, \Delta }
 \end{prooftree} \\[9 pt]
  \end{tabular}
\end{minipage}\begin{minipage}[t]{0.5\textwidth}\begin{tabular}{rl}
  \dls{impliesR} & \begin{prooftree}
  \hypo{\Gamma, \, P \vdash Q, \, \Delta }
  \infer1
    {\Gamma \vdash  P \rightarrow Q, \,  \Delta }
 \end{prooftree} \\[9 pt]
   \dls{impliesL} & \begin{prooftree}
  \hypo{\Gamma \vdash P, \, \Delta \,\,\, & \Gamma, \, Q \vdash \Delta}
  \infer1
    {\Gamma,\, P \rightarrow Q \vdash \Delta }
 \end{prooftree} \\[9 pt]
  \dls{iffR} & \begin{prooftree}
  \hypo{\Gamma, P \vdash Q,  \Delta  \,\,\,  &  \Gamma, Q \vdash P, \Delta}
  \infer1
    {\Gamma \vdash P \leftrightarrow Q, \,  \Delta }
 \end{prooftree} \\[9 pt]
 \dls{iffL} & \begin{prooftree}
  \hypo{\Gamma, P \wedge Q \vdash \Delta  \,\,\,  &  \Gamma, \neg P \wedge \neg Q \vdash \Delta}
  \infer1
    {\Gamma, P \leftrightarrow Q \vdash \Delta }
 \end{prooftree} \\[9 pt]
       \dls{falseL} & \begin{prooftree}
  \hypo{}
  \infer1
    {\Gamma, \, \bot \vdash \Delta }
 \end{prooftree} \\[9 pt]
       \dls{trueR} & \begin{prooftree}
  \hypo{}
  \infer1
    {\Gamma \vdash \top, \, \Delta }
 \end{prooftree} \\[9 pt]
    \dls{axiom} & \begin{prooftree}
  \hypo{}
  \infer1
    { \Gamma, \, P  \vdash P, \, \Delta}
 \end{prooftree} \\[9 pt]
   \dls{weakL} & \begin{prooftree}
  \hypo{P, \Gamma \vdash \Delta \, \, \,& Q \vdash P}
  \infer1
    {\Gamma, Q  \vdash \Delta }
 \end{prooftree} \\[9 pt]
\end{tabular}
\end{minipage}}
} 
\caption{Propositional \DDL{} rules} \label{fig:proprewrites} \end{figure}

  \begin{figure}[t]
  
\noindent\ovalbox{
{ \small
\begin{minipage}[t]{0.95\textwidth}
\begin{tabular}{rl} \dls{existsR} & \begin{prooftree}
  \hypo{\Gamma \vdash  p(e), \, \Delta}
  \infer1[(any $e$)]
    {\Gamma  \vdash \exists x : p(x), \,  \Delta}
 \end{prooftree} \\[9 pt]
   \dls{forallL} & \begin{prooftree}
  \hypo{\Gamma, \, p(e) \vdash \Delta}
  \infer1[(any $e$)]
    {\Gamma, \, \forall x  : p(x)  \vdash \Delta}
 \end{prooftree} \\[9 pt]
 \dls{forallR} & \begin{prooftree}
  \hypo{\Gamma \vdash p(y), \, \Delta}
  \infer1[($y$ Skolem symbol)]
    {\Gamma \vdash  \forall x : p(x), \Delta}
 \end{prooftree} \\[9 pt]
   \dls{existsL} & \begin{prooftree}
  \hypo{\Gamma, \, p(y) \vdash \Delta}
  \infer1[($y$ Skolem symbol)]
    {\Gamma, \,\exists x : p(x)  \vdash \Delta}
 \end{prooftree} \\[9 pt]
\end{tabular}
\end{minipage}}
} 
\caption{Quantification \DDL{} rules} \label{fig:quantrules} \end{figure}

  \begin{figure}[t]
  
\noindent\ovalbox{
  { \small
\begin{minipage}[t]{0.45\textwidth}
\begin{tabular}{rl}
\dls{moveR} & \begin{prooftree}
  \hypo{ \Gamma \vdash Q, \, P, \, \Delta }
  \infer1
    {  \Gamma \vdash P, \, Q, \, \Delta}
 \end{prooftree} \\[9 pt]
 \dls{moveL} & \begin{prooftree}
  \hypo{ \Gamma, \, Q, \, P \vdash \Delta }
  \infer1
    {  \Gamma,\, P, \, Q \vdash  \Delta}
 \end{prooftree} \\[9 pt]
\end{tabular}
\end{minipage}\begin{minipage}[t]{0.5\textwidth}\begin{tabular}{rl}
      \dls{hideR} & \begin{prooftree}
  \hypo{\Gamma  \vdash \Delta}
  \infer1
    {\Gamma \vdash P, \, \Delta }
 \end{prooftree} \\[9 pt]
     \dls{hideL} & \begin{prooftree}
  \hypo{\Gamma  \vdash \Delta}
  \infer1
    {\Gamma, \, P \vdash \Delta }
 \end{prooftree} \\[9 pt]
\end{tabular}
\end{minipage}
}
} 
\caption{Structural \DDL{} rules} \label{fig:structrules} \end{figure}
The propositional rules in \DDL{} 
allow manipulation of the basic logical connectives ($\wedge$, $\vee$, $\neg$,
$\rightarrow$, $\iff$) and operators ($\top$,
$\bot$) in the \DDL{}-sequent (see Figure \ref{fig:proprewrites}). 
For
example, the rule \dlref{impliesR}, defined as
\begin{center} 
\begin{tabular}{cc}  &  \begin{prooftree}
  \hypo{\Gamma, \, P \vdash Q, \, \Delta }
  \infer1
    {\Gamma \vdash  P \rightarrow Q, \,  \Delta,}
 \end{prooftree} \end{tabular} \end{center} 
allows an implication in the \DDL{}-consequent, $P \rightarrow Q$ to be simplified to $P$ in the \DDL{}-antecedent and $Q$ in the \DDL{}-consequent. 
  Here, $\Gamma \vdash  P
\rightarrow Q, \,  \Delta$ is the \DDL{}-sequent that \dlref{impliesR}
can be applied to and $\Gamma, \, P \vdash Q, \, \Delta$ is the
simplified \DDL{}-sequent. Note that the standard logical notation being used for \dlref{impliesR} above is for ease of presentation, whereas the PVS specification of such a rule, generally hidden from a user by a strategy, is closer to that described in Section \ref{sec:seq}. Additionally, there are quantification
rules for Skolemization and instantiation in the \DDL{}-sequent (see Figure \ref{fig:quantrules}) and there are structural rules that allow expressions to be moved or deleted (Figure \ref{fig:structrules}).

In addition to the propositional, quantification, and structural rules,
\name{} provides a collection of powerful
proof commands that combine the more basic \DDL{} strategies.
A list these
additional proof commands is given in Figure~\ref{fig:batch}.

\begin{figure}[t]
\noindent\ovalbox{
{ \small \begin{minipage}[t]{\textwidth}
\vspace{5 pt}
 \begin{itemize}[align=parleft,leftmargin=*,labelindent=0mm,labelsep=15mm]
  \item[~\mbox{\dll{dl-flatten}}]  Disjunctively simplifies the \DDL{} sequent by applying \dlref{trueR}, \dlref{falseL}, \dlref{orR}, \dlref{impliesR}, \dlref{notR}, \dlref{axiom}, \dlref{falseL}. 
    \item[~\mbox{\dll{dl-ground}}]  Disjunctively and conjunctively simplifies the \DDL{} sequent by applying \dlref{dl-flatten} and additional splitting lemmas \dlref{andR}, \dlref{orL}, and \dlref{impliesL}. 
        \item[~\mbox{\dll{dl-inst}}]  Instantiates a universal quantifier in the  \DDL{}-antecedent by applying \dlref{forallL} or an existential quantifier in the \DDL{}-consequent by applying \dlref{existsL}.
            \item[~\mbox{\dll{dl-skolem}}]  Skolemizes an existential quantifier in \DDL{}-antecedent by applying \dlref{existsR} or a universal quantifier in the \DDL{}-consequent by applying \dlref{forallR}. 
            \item[~\mbox{\dll{dl-grind}}] Repeatedly uses
              \dlref{dl-ground} and \dll{skolem} and serveral
              rewriting rules related to real expressions. This strategy has the option to use the MetiTarski automatic theorem prover as an outside oracle to discharge the proof if possible. 
  \item[~\mbox{\dll{dl-assert}}] Repeatedly applies hybrid program
    rewriting rules in Figure \ref{fig:hprewrites}.
 \end{itemize}
\end{minipage}}}
\caption{\DDL{} proof commands} \label{fig:batch}
\end{figure}
 
\begin{example}[\DDL{}-sequent example] \label{ex:dlseq1}
The \DDL{}-sequent 
 $$  \vdash (x = c \wedge y = 0) \rightarrow
 \left[\alpha\right] \pathc(c).$$
expresses the validity of the expression in Formula~\ref{eqn:ale} from 
Example~\ref{ex:ale}. Invoking the rule \dlref{dl-flatten} to the sequent above applies
 \dlref{impliesR} and \dlref{andL}, which separates conjunctions in
 the antecedent, resulting in the following \DDL{}-sequent:
  \begin{align} \label{eq:seq1} x = c, \, y = 0  \vdash  \left[\alpha\right] \pathc(c). \end{align}
\end{example}

  \subsection{Hybrid program rewriting rules} \label{sec:hprr}
  While the rules in Section \ref{ref:log} manipulate the logical structure of a \DDL{}-sequent, further rules act on the hybrid program components of such a sequent. Properties given in Figure \ref{fig:hprewrites} allow direct rewriting of hybrid programs. Other
  rules about hybrid programs in a sequent are given in Figure
  \ref{fig:hprules}. Most of these rules manipulate the allruns $\left[\, \cdot \, \right]$
  or someruns $\left\langle \, \cdot \, \right\rangle$ operators and the proofs were largely concerned with reasoning about the
  semantic relation function \sr{} defined in Section~\ref{sec:HP}.
In addition to each of these rules becoming strategies, the 
command \dlref{dl-assert} uses all the hybrid program rewriting rules in Table \ref{fig:hprewrites} to simplify an expression. 

  \begin{figure}[t]
\noindent\ovalbox{
{ \small
\begin{minipage}[t]{0.95\textwidth}
\begin{tabular}{rl}
\dls{boxd}  &  $\langle \alpha \rangle P \leftrightarrow \neg \left[\alpha\right]\neg P$  \\[3 pt]
\dls{assignb}  &   $\left[\mathbf{x} := \ell\right]P = \SUB{}(\mathbf{x} := \ell)(P) $ \\[3 pt]
\dls{assignd}  &   $\left\langle \mathbf{x} := \ell\right\rangle P = \SUB{}(\mathbf{x} := \ell)(P) $ \\[3 pt]
\dls{testb} & $\left[?Q\right] P = Q \rightarrow P$ \\[3 pt]
\dls{testd} & $\left\langle ?Q \right\rangle P = Q \wedge P$ \\[3 pt]
\dls{choiceb} & $\left[\alpha_{1} \cup \alpha_{2}\right]P \leftrightarrow \left[\alpha_{1}\right]P \wedge \left[\alpha_{2}\right]P$ \\[3 pt]
\dls{choiced} & $\left\langle\alpha_{1} \cup \alpha_{2}\right\rangle P \leftrightarrow \left\langle\alpha_{1}\right\rangle P \vee \left\langle \alpha_{2}\right\rangle P$ \\[3 pt]
\dls{composeb} & $\left[\alpha_{1};\alpha_{2}\right]P \leftrightarrow \left[\alpha_{1}\right]\left[\alpha_{2}\right]P$ \\[3 pt] 
\dls{composed} & $\left\langle\alpha_{1};\alpha_{2}\right\rangle P \leftrightarrow \left\langle \alpha_{1}\right\rangle \left\langle \alpha_{2}\right\rangle P$ \\[3 pt] 
\dls{iterateb} & $\left[\alpha^{*}\right]P = P \wedge \left[\alpha\right]\left[\alpha^{*}\right]P$ \\[3 pt]
\dls{iterated} & $\left\langle\alpha^{*}\right\rangle P = P \vee \left\langle \alpha\right\rangle \left\langle\alpha^{*}\right\rangle P$ \\[3 pt]
\dls{anyb} &  $\left[x := * \right]P(x) = \forall x : P(x)$  \\[3 pt]
\dls{anyd}  &  $\left\langle x := * \right\rangle P(x) = \exists x : P(x)$  \\[3 pt]
\end{tabular}
\end{minipage}}
}
\caption{Hybrid program rewriting rules.} \label{fig:hprewrites} \end{figure}
 \begin{figure}[t]
\noindent\ovalbox{
{ \small \begin{minipage}[t]{0.45\textwidth}
\begin{tabular}{rl}
\dls{Mb} & \begin{prooftree}
  \hypo{  \vdash P \rightarrow Q }
  \infer1
    { \Gamma \vdash \left[\alpha\right]P \rightarrow \left[\alpha\right]Q,  \Delta }
 \end{prooftree} \\[9 pt]
 \dls{Md} & \begin{prooftree}
  \hypo{  \vdash P \rightarrow Q }
  \infer1
    { \Gamma \vdash \left\langle \alpha \right\rangle P \rightarrow \left\langle \alpha \right\rangle Q,  \Delta }
 \end{prooftree} \\[9 pt]
\dls{K} & \begin{prooftree}
  \hypo{ \Gamma \vdash \left[\alpha\right](P \rightarrow Q), \Delta }
  \infer1
    { \Gamma \vdash \left[\alpha\right]P \rightarrow \left[\alpha\right]Q,  \Delta }
 \end{prooftree} \\[9 pt]
 \dls{loop} & \begin{prooftree}
  \hypo{ \Gamma \vdash J, \Delta \,\,\, & J \vdash \left[\alpha\right]J \,\,\,\,\,\,  J \vdash P }
  \infer1
    { \Gamma \vdash \left[\alpha^{*}\right]P, \Delta }
 \end{prooftree} \\[9 pt]
  \dls{mbR} & \begin{prooftree}
  \hypo{ \Gamma \vdash \left[\alpha\right]Q, \Delta \,\,\, & Q \vdash P }
  \infer1
    { \Gamma \vdash \left[\alpha\right]P, \Delta }
 \end{prooftree} \\[9 pt]
  \dls{mbL} & \begin{prooftree}
  \hypo{ \Gamma, \left[\alpha\right]Q \vdash \Delta \,\,\, & P \vdash Q }
  \infer1
    { \Gamma, \left[\alpha\right]P \vdash \Delta}
 \end{prooftree} \\[9 pt]
  \dls{ghost} & \begin{prooftree}
  \hypo{ \Gamma \vdash \left[y := e \right]P, \Delta}
  \infer1[$\fresh(P)(y)$]
    { \Gamma \vdash P, \Delta}
 \end{prooftree} \\[9 pt]
   \end{tabular}
\end{minipage}\begin{minipage}[t]{0.5\textwidth}\begin{tabular}{rl}
  \dls{Gb} & \begin{prooftree}
  \hypo{\vdash P  }
  \infer1[]
    {\Gamma \vdash \left[\alpha\right] P, \Delta }
 \end{prooftree} \\[9 pt]
 \dls{Gd} & \begin{prooftree}
  \hypo{\vdash \langle \alpha \rangle \bm{\top} \, \, \,\vdash P  }
  \infer1[]
    {\Gamma \vdash \left\langle \alpha\right\rangle P, \Delta }
 \end{prooftree} \\[9 pt]
  \dls{VRb} & \begin{prooftree}
  \hypo{\Gamma \vdash P, \Delta }
  \infer1[{\footnotesize $\fresh(P)(\alpha)$  } ]
    {\Gamma \vdash   \left[\alpha\right]P, \Delta }
 \end{prooftree} \\[9 pt]
   \dls{VRd} & \begin{prooftree}
  \hypo{\vdash \langle \alpha \rangle \bm{\top}  \, \, \, \Gamma \vdash P, \Delta }
  \infer1[{\footnotesize $\fresh(P)(\alpha)$  } ]
    {\Gamma \vdash   \left\langle \alpha\right\rangle P, \Delta }
 \end{prooftree} \\[9 pt]
   \dls{mdR} & \begin{prooftree}
  \hypo{ \Gamma \vdash \left\langle \alpha\right\rangle Q, \Delta \,\,\, & Q \vdash P }
  \infer1
    { \Gamma \vdash \left\langle \alpha\right\rangle P, \Delta }
 \end{prooftree} \\[9 pt]
  \dls{mdL} & \begin{prooftree}
  \hypo{ \Gamma, \left\langle \alpha\right\rangle Q \vdash \Delta \,\,\, & P \vdash Q }
  \infer1
    { \Gamma, \left\langle \alpha\right\rangle P \vdash \Delta}
 \end{prooftree} \\[9 pt]
\end{tabular}
\end{minipage}}
}
\caption{Hybrid program rules.} \label{fig:hprules} \end{figure}

There are a few intricacies worth mentioning in the formal
verification and implementation of these rules in PVS.
In the rewriting rules \dlref{assignb} and \dlref{assignd}, an allruns or
someruns of an assignment HP is equated to a substitution. 
Substitution is defined at the environment level as follows.
\begin{align} \asub(\mathbf{x} := \ell)(e)(i) \ \dfn\ \begin{cases}
\ell(i)(e) &  \textrm{if } i \in \mathbf{x}   \\
e(i)   & \textrm{if } i \not \in \mathbf{x}.
\end{cases} \end{align} Substitution of a general Boolean expression is therefore defined as follows.
$$ \SUB(\mathbf{x} := \ell)(P) \ \dfn\ \lambda(e: \mathcal{E}). P(\asub(\mathbf{x} := \ell)(e)).$$
While the definition of substitution above applies to any Boolean
expression $P$ and can be reasoned about by a user of Plaidypvs, the
standard level of manipulation in \DDL{} is not often at the
environment level. To increase the level of automation, several
rewriting rules for reducing expressions containing $\SUB$ have been implemented. 
This led to formally verifying substitution properties for real expressions, inequalities of real
expressions, and hybrid programs, so that a substitution at the top
level of an expression could be pushed down to the level of $\val$ and
$\cnst,$ where atomic substitutions are applied. The implementation of these rules required a calculus for reducing the substitution down to atomic expressions, written in the strategy language of PVS.
This allows the \dlref{assignb} and \dlref{assignd} strategies to automatically compute a substitution for any propositional expression composed of equalities and inequalities of polynomial real expressions. For example, the substitution
\begin{align*} \SUB( x := y, y := 10)&(x^2 + y^2 = 11),\end{align*}
is transformed automatically into $y^2 + 10^2 = 11$ as follows.
\begin{align*}
 \SUB( x := y, y := 10)(x^2 + y^2 = 11)  & = \left( \SUBre( x := y, y := 10)(x^2 + y^2) =  \SUBre( x := y, y := 10)(11) \right)\\ 
 & =  \left(\SUBre( x := y, y := 10)(x^2) +  \SUBre( x := y, y := 10)(y^2) =  11 \right)\\
  & =  \left(\SUBre( x := y, y := 10)(x)^2 +  \SUBre( x := y, y := 10)(y)^2 =  11 \right)\\
 & =   y^2 + 10^2 = 11,
\end{align*} where \SUBre{} is substitution defined on real expressions $r \in \mathcal{R}$ as
$$ \SUBre(\mathbf{x} := \ell)(r) \ \dfn\  \lambda(e: \mathcal{E}).  r(\asub(\mathbf{x} := \ell)(e)).$$
The automated substitution of more general Boolean expressions (for example, a statement of the form $\left[\alpha\right]P$) is still incomplete in Plaidypvs, and an area of future work. 

Another challenge in formal verification occurs in some hybrid
program rules. The \dlref{ghost}, \dlref{VRb}, and \dlref{VRd} rules require the concept of \emph{freshness}. A fresh variable $y$ is defined as 
\begin{align*}
\fresh(P)(y) \ \dfn\ \forall e \in \mathcal{E}, \, r \in \mathbb{R}, \, P(e) = P(e \textrm{ with } y \mapsto r)].
\end{align*}
In other words, the value of the Boolean expression $P$ does not
depend on the value of the variable $y$. Analogous definitions exist
to express that a variable is fresh relative to a real expression or a
hybrid program. Furthermore, an entire hybrid program can be checked
for freshness relative to a Boolean expression as follows.
\begin{align} \fresh(P)(\alpha) \ \dfn\ \begin{cases}
\forall k \in \mathbf{x}\, \,\,\, \fresh(P)(k) & \textrm{if } \alpha = (\mathbf{x} := \ell),   \\
\forall k \in \mathbf{x}'\, \,\,\, \fresh(P)(k) & \textrm{if } \alpha = (\mathbf{x}' = \ell\, \& \,Q), \\ 
\True & \textrm{if } \alpha = ?Q, \\ 
\fresh(P)(x)  & \textrm{if } \alpha = (x := * \, \, \& \, Q), \\ 
\fresh(P)(\alpha_1) \wedge \, \fresh(P)(\alpha_2)  & \textrm{if } \alpha = \alpha_1 ;\alpha_2,  \\
\fresh(P)(\alpha_1) \wedge \, \fresh(P)(\alpha_2)  & \textrm{if } \alpha =  \alpha_1 \cup \alpha_2, \\
 \fresh(P)(\alpha)   & \textrm{if } \alpha =  \alpha_{1}^{*}.  \\
 \end{cases} \end{align}
 Note that the recursive definition of freshness above ensures the value of $P$ does not change for any run of the hybrid program $\alpha$ by checking if all the variables potentially changing in $\alpha$ are fresh relative to $P$. 
 \vspace{-13 pt}
 
The need for a fresh variable, as in the rule \dlref{ghost}, requires a mechanism for producing
fresh variables relative to a \DDL{}-sequent. Since variables are represented by indices, a fresh
variable can be generated by computing the smallest natural number in a \DDL{}-sequent
not being used as a variable index. \name{} also provides strategies
for
automatically proving freshness of variables in \DDL{}-sequents.
\begin{example}[\DDL{}-sequent example continued] \label{ex:dlseq2}
Expanding $\alpha$ in the sequent given by Formula~\ref{eq:seq1}, from
Example \ref{ex:dlseq1}, and using \dlref{loop} with $J =(\pathc(c) \wedge y \geq 0)$ produces three subgoals,\footnote{The other
 two \DDL{}-sequents generated can be proven easily. For full details of the examples in this paper, see the PVS implementation at \href{https://github.com/nasa/pvslib/tree/master/dL/examples}{https://github.com/nasa/pvslib/tree/master/dL/examples}} one of which is 
\begin{align*}  \pathc(c),  y \geq 0  \vdash & \left[ (?(x > 0); (x' = -y, y' = x, \& \, x \geq 0)) \, \cup \right.  \\
 & \left. (?(x \leq 0); (x' = -c, y' = 0)) \right] \pathc(c) \wedge y \geq 0 . \end{align*}
 Using \dlref{dl-assert} to simplify with hybrid program rewriting rules and applying propositional simplifications with the command \dlref{dl-ground} result in the following two \DDL{}-sequents
\begin{align} \label{eq:seq2} \begin{split} (x > 0, \circc(c), y \geq 0 ) & \vdash  \left[ x' = -y, y' =
                                        x, \& \, x \geq 0))  \right]
                                        \pathc(c) \wedge y \geq 0, \\ (x \leq 0,  y = c)  &\vdash  \left[ (x' = -c, y' = 0)  \right]  \pathc(c) \wedge y \geq 0. \end{split} \end{align}
 \end{example}
                                  
\begin{figure}
\noindent\ovalbox{
{ \small \begin{minipage}[t]{0.95\textwidth}
\begin{tabular}{rl}
\dls{dinit} & \begin{prooftree}
  \hypo{ \Gamma, Q \vdash \left[x' = f(x) \, \& \, Q \right]P, \Delta }
  \infer1
    { \Gamma \vdash \left[x' = f(x) \, \& \, Q \right]P, \Delta}
 \end{prooftree} \\[12 pt]
\dls{dW} & \begin{prooftree}
  \hypo{ Q \vdash P }
  \infer1
    { \Gamma \vdash \left[ x' = f(x) \, \& \, Q \right]P, \Delta  }
 \end{prooftree}  \\[12 pt]
 \dls{dI} & \begin{prooftree}
  \hypo{\Gamma, Q \vdash P, \Delta \, \, \, Q \vdash \left[x' := f(x) \right](P)' }
  \infer1
    { \Gamma \vdash \left[x' = f(x) \, \& \, Q\right]P, \Delta  }
 \end{prooftree}  \\[12 pt]
  \dls{dC} & \begin{prooftree}
  \hypo{ \Gamma \vdash \left[x' = f(x) \, \& \, Q \right]C, \Delta \,\,\, & \Gamma \vdash \left[x' = f(x) \, \& \, (Q \wedge C) \right]P, \Delta }
  \infer1
    { \Gamma \vdash \left[x' = f(x)\, \&\, Q\right]P, \Delta  }
 \end{prooftree} \\[12 pt]
   \dls{dG} & \begin{prooftree}
  \hypo{\Gamma \vdash G, \, G \vdash P, \,  \Gamma \vdash \exists y \left[x' = f(x),\, y' = a(x) \cdot y + b(x) \, \& \, Q\right]G, \Delta  }
    \infer1[{\footnotesize $\fresh(y)$  } ]
    {\Gamma \vdash \left[x' = f(x)\, \&\, Q\right]P, \Delta  }
 \end{prooftree}  \\[12 pt]
     \dls{dS} & \begin{prooftree}
  \hypo{\Gamma \vdash \forall t \geq 0 \left(\forall 0 \leq s \leq t\,  Q(y(s))\right) \rightarrow  \left[x := y(t)\right] P   }
  \infer1
    {\Gamma \vdash \left[x' = f(x) \, \& \, Q\right]P }
 \end{prooftree} \\[9 pt]    \end{tabular}
    \end{minipage}
}
}

\caption{Differential Equation Rules. For \protect\dlref{dG}, $a$ and $b$ are continuous on $Q$, and $y$ is fresh relative to $x' = f(x)$, $Q$, $a$, $b$, $P$, $\Gamma$ and $\Delta$.}\label{fig:diffrules} \end{figure}
\subsection{Rules for differential equations}
   The rules for differential equations are given in Figure
   \ref{fig:diffrules}. The differential equation rules required significant mathematical underpinnings to be added to PVS for their formal verification. 
 For the implementation of the
\dlref{dI} rule, a calculus to automatically compute the derivative of a
Boolean expression $P$ was necessary. To do this, an embedding of
non-quantified Boolean expressions was developed as a data type with
the following grammar.
\begin{align*}  b ::=\ &  b_{1} \wedge_{\nqB} b_{2} \  \rvert\ b_{1} \vee_{\nqB} b_{2}\ \rvert\  \neg_{\nqB}b_{1}\ \rvert\ \rel_{\nqB}(r_{1},r_{2}),  \end{align*}
where $\rel_{\nqB}$ is of type $\NQBr$, which is itself an embedding of the following inequality operators.
\begin{align*}   \rel_{\nqB} ::=\ & \leq_{\nqB}\ \rvert\ \geq_{\nqB}\ \rvert\ <_{\nqB}\ \rvert\ >_{\nqB}\ \rvert\ =_{\nqB}\ \rvert\  \neq_{\nqB}.\ \end{align*}
With this structure, the derivative $b'$ of a Boolean expression $b$
is defined as \[b' \ \dfn\ \begin{cases}
b_{1}' \wedge b_{2}' & \textrm{if } 
b = b_{1} \wedge_{\nqB} b_{2}  \textrm{ or } 
b = b_{1} \vee_{\nqB} b_{2}   \\ 
r'_{1} \leq r'_{2} & \textrm{if } 
b = r_{1} \leq_{\nqB} r_{2} 
 \textrm{ or } 
b = r_{1} <_{\nqB} r_{2}   \\ 
r'_{1} \geq r'_{2} & \textrm{if } 
b = r_{1} \geq_{\nqB} r_{2}  \textrm{ or } 
b = r_{1} >_{\nqB} r_{2}   \\ 
r'_{1} = r'_{2} & \textrm{if } 
b = \left(r_{1} =_{\nqB} r_{2}\right)   \textrm{ or } 
b = \left(r_{1} \neq_{\nqB} r_{2} \right). 
\end{cases} \]
In PVS, $\left[x' := f(x) \right](P)'$ is computed by replacing $P$
with its equivalent non-quantified Boolean, and the derivative of any real expression $r$ occurring in $P$ is the real expression given by:
$$r' = \sum_{i\in \mathbf{x}}\partial r_{i} \cdot \ell(i).$$
This is the derivative of the real expression $r$ in terms of the explicit variable that all the variables in $\mathbf{x}$ are a function of. To arrive at this formulation, differentiability and partial differentiability had to be defined for real expressions as well as the multivariate chain rule.

For the Differential Ghost rule \dlref{dG}, adding an equation to the
differential equation $x' = \ell$ required that the new differential
equation $x' = \ell, y' = a(x)\cdot y + b(x)$ had a unique
solution. The Picard-Lindel\"{o}ff theorem can be used to show that if $a$
and $b$ are continuous on $Q$, then there is a unique solution to $y'
= a(x) \cdot y + b(x)$. Given a solution to $x'= \ell$ that is
contained in $Q$, it follows that $x' = \ell, y' = a(x)\cdot y + b(x)$
has a unique solution. These properties of differential equations,
including the Picard-Lindel\"{o}ff theorem, were developed in PVS specifically
to prove these rules.

\begin{example}[\DDL{}-sequent example continued] \label{ex:dlseq3}
Applying \dlref{dC} with $C = \circc(c)$ to the first branch of the proof in Example \ref{ex:dlseq2}, eq. \ref{eq:seq2} produces two subgoals, the first of which (with expanded $\circc$) is 
 \begin{align*} (x > 0, x^2 + y^2 = c^2, y \geq 0)  \vdash  \left[ x' = -y, y' = x \& \, x \geq 0))  \right]
 (x^2 + y^2 = & c^2).  \end{align*}
 \end{example}
 Using $\dlref{dI}$ reduces to two cases:
\begin{align} \label{eq:seq22} \begin{split} x \geq 0   & \vdash 
 (2\cdot x \cdot -y + 2 \cdot y \cdot x =  0)  \\ (x \geq 0, x^2 + y^2 = c^2, y \geq 0)  &\vdash  x^2 + y^2 = c^2,  \end{split} \end{align}
both of which can be proven with basic algebraic and logical simplifications included in command $\dlref{dl-grind}$.

 \section{Using \name{}} \label{sec:using}
\begin{figure}[t]
\begin{center}
\includegraphics[width=.75\textwidth]{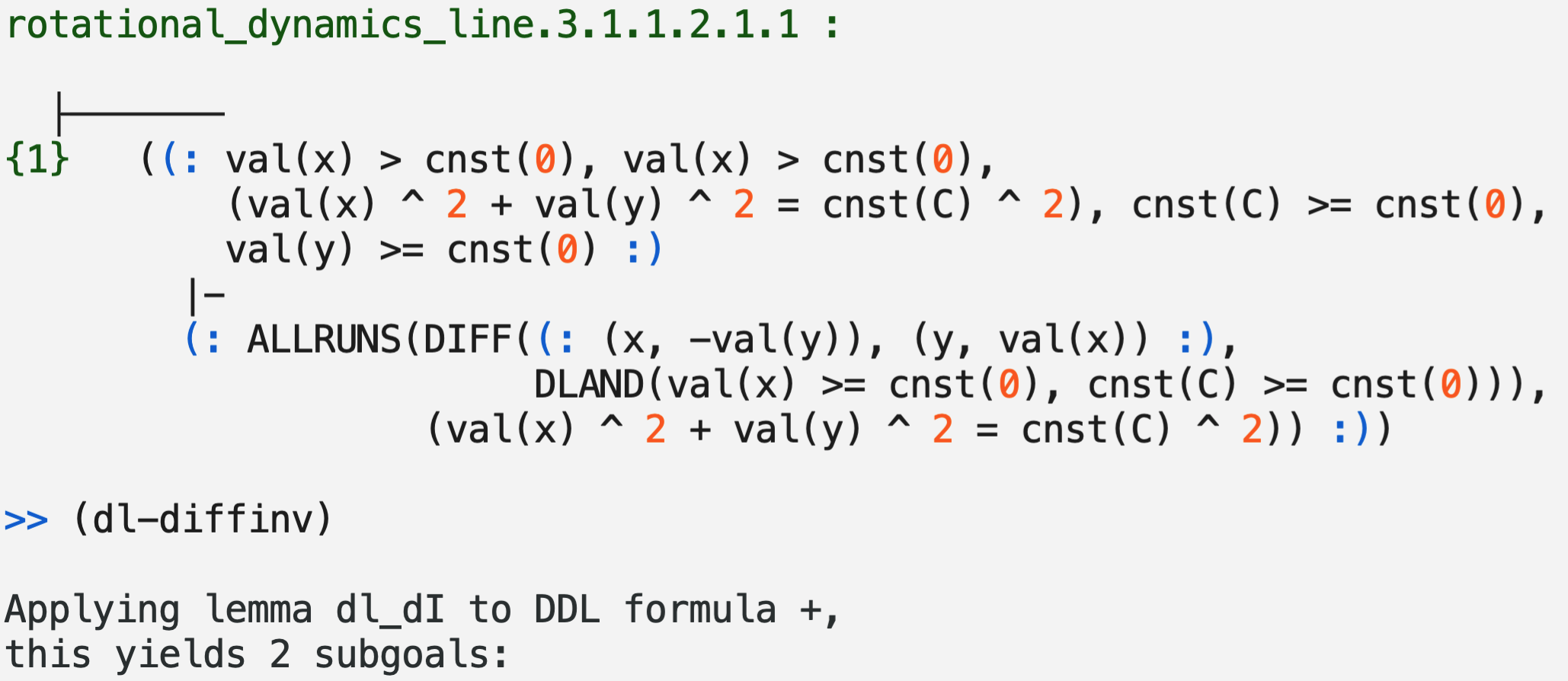}
\includegraphics[width=.75\textwidth]{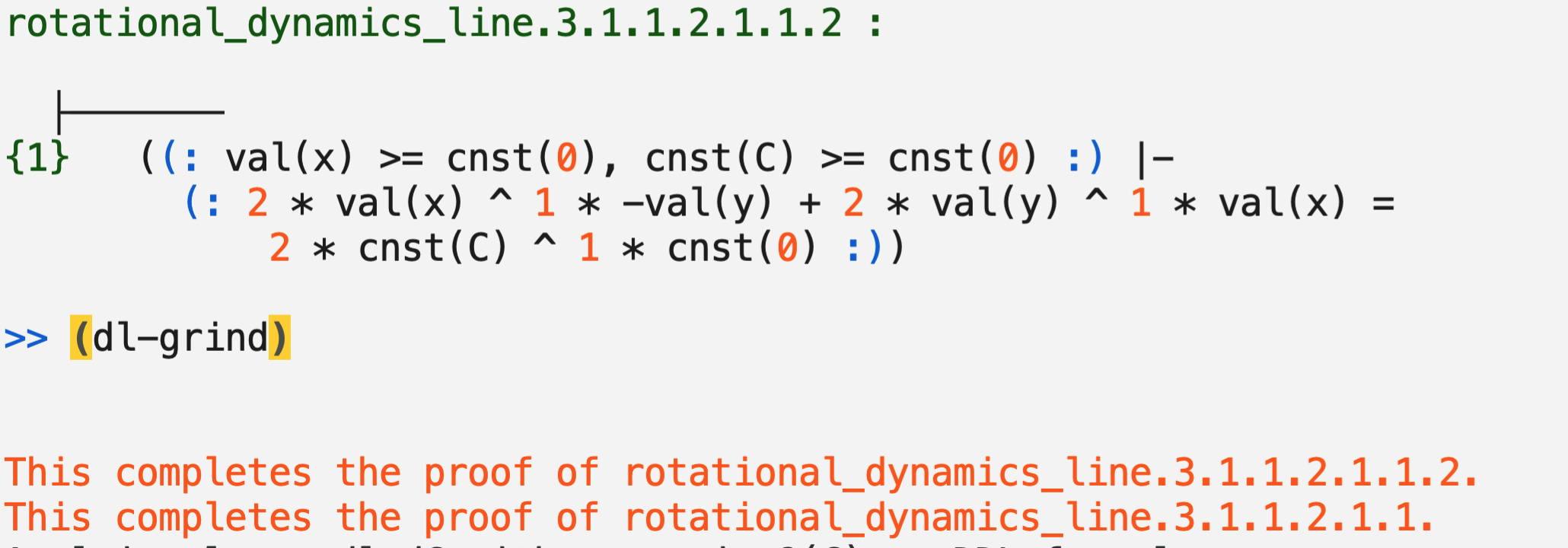}
\end{center}
\caption{The proof steps that complete the proof discussed in Example \ref{ex:dlseq3}.} \label{fig:screenshot23}
\end{figure}

\name{} provides the functionality of \DDL{} within the PVS environment. Numerous examples can be found in the directory \texttt{examples} of the \name{} library. 
\begin{figure}[t]
\begin{center}
\includegraphics[width=.75\textwidth]{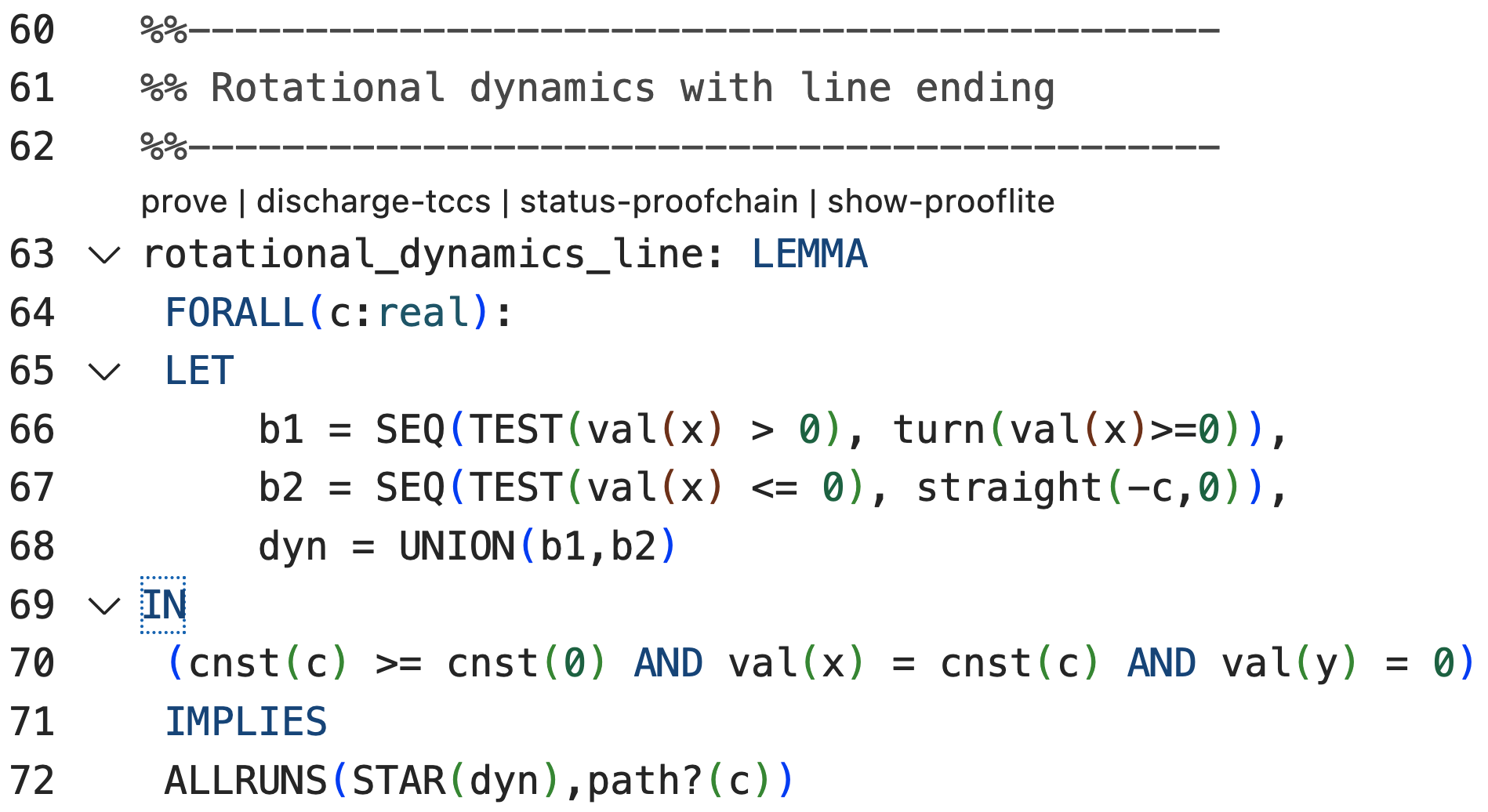}
\end{center}
\caption{The specification of Example \ref{ex:ale} in \name.}\label{fig:screenshot1}
\end{figure}
Figures \ref{fig:screenshot1} and \ref{fig:screenshot23} illustrate using \DDL{} for specification and verification of hybrid systems in \name. However, \name{} is not limited to just these applications, the embedding allows additional features to be used for formal reasoning of hybrid programs. For example, the definition of other functions from PVS libraries can be imported into a formal development that uses \name{}. Furthermore, meta-properties about hybrid programs can be specified and proven.
The example below illustrates these features. 

\begin{example}[Verified connection to Dubins paths] \label{ex:dub} 
An aircraft moving at a constant speed $c > 0$ with a turn rate of $1$ can be modeled by a Dubins path:
$$ \theta' = 1, x' = -c\sin(\theta), y' = c\cos(\theta).$$ Furthermore, it can be shown that the hybrid program $\beta$ defined as
\begin{align*} ((&?(x \geq 0); ( \theta' = 1, x' = - c \sin(\theta), y' = c \cos(\theta)\& \, x \geq 0)) \, \cup  \\ 
& (?(x < 0); (x' = -c, y' = 0)))^{*}, \end{align*}
is equivalent to the hybrid program $\alpha$ defined in Example \ref{ex:hpe}, for appropriate initial values. Formally, this is a property relating the $\sr$ function associated with each of these programs, namely for environments $e_{i}, e_{o}$ such that $e_{i}(x) = c$ and $e_{i}(y) = 0$
\begin{align*}
\left(\exists t: \, \,  \sr(\beta)(e_{i} \textrm{ with } [\theta := 0])(e_{o}\textrm{ with }\right.&\left.[\theta := t])\right)  \iff \\ & \sr(\alpha)(e_{i})(e_{o} \textrm{ with } [\theta := e_{i}(\theta)]). \end{align*}
Note the property above involves generic hybrid programs rather than particular instances. Thus, for a Boolean expression  $Q$ that does not change according to $\theta$:
$$(x=c,y=0  \rightarrow \left[\alpha\right]Q) \iff (x=c,y=0,\theta=0  \rightarrow \left[\beta\right]Q).$$
\end{example}

  \section{Related work} \label{sec:related}
There is a long line of research on the formal verification of hybrid systems. The development of \DDL{} itself (\cite{DBLP:conf/tableaux/Platzer07,platzer2008differential,platzer2017complete,Platzer18}) and its use in formal verification of hybrid systems (\cite{DBLP:journals/ral/BohrerTMSP19,DBLP:journals/tecs/CleavelandMP23,DBLP:conf/aaai/FultonP18,DBLP:journals/sttt/JeanninGKSGMP17,KabraMP22,DBLP:conf/rssrail/MitschGBGP17,DBLP:journals/ijrr/MitschGVP17,DBLP:conf/fase/MullerMRSP17}) is well-known. Additionally, there has been significant work done in the PVS theorem prover~\cite{abraham2001verification,slagel2021formal},  Event-B~\cite{dupont2021correct}, and Isabelle/HOL~\cite{foster2021hybrid,huerta2020algebraic,huerta2018verifying,huerta2022predicate,sheng2023hhlpy,struth2021hybrid,wang2015improved,zou2015formal} verifying hybrid systems outside of the \DDL{} framework. 

The most similar verification effort to the current development is~\cite{bohrer2017formally}, where the authors formally verified the soundness of \DDL{} in Coq and Isabelle. The work in~\cite{bohrer2017formally} focuses on a full formal verification of soundness of  \DDL{}, with the goal of a formally verified prover kernel for KeYmaera X. The result are proof checkers in Coq and Isabelle for \DDL{} proofs. The goal of \name{} is a verified  operational embedding of \DDL{} in the theorem prover PVS, allowing specification and reasoning about HPs \emph{interactively} within PVS. 

While the work in~\cite{bohrer2017formally} proves soundness of most of the proof calculus of \DDL{}, the present work focuses on verifying the proof \emph{rules} of \DDL{}.  Particularly, the substitution axiom in \DDL{} that allows rules and axioms to be applied to specifications of HPs in \DDL{} is proven in~\cite{bohrer2017formally} but not directly proven for the PVS embedding. Instead, substitution is handled by the instantiation functionalities of PVS itself, specifically when \DDL{} rules and axioms are applied as strategies to a particular \DDL{}-sequent in the interactive prover.  Additionally, there are several places where the embedding of \DDL{} in this work is more general than the work in~\cite{bohrer2017formally}. 
Differential Ghost and Differential Effect in ~\cite{bohrer2017formally} are shown for a single ordinary differential equation rather than the more general \emph{system} of ordinary differential equations. Differential Solve is only shown for differential equations with linear solutions, whereas the corresponding rule in Plaidypvs automatically solves differential equations with linear and quadratic solutions and is proven for any ODE where the solution is known. Differential Invariant in~\cite{bohrer2017formally} is restricted to propositions of the form $P = (f(x) \geq g(x))$ and $P = (f(x) > g(x))$ and it is remarked that other cases can be derived in \DDL{} from these two cases, but in Plaidypvs Differential Invariant is fully implemented for any proposition that is the conjunction or disjunction of inequalities.

The current work formalizes a version of \DDL{} based on Parts I and II in~\cite{Platzer18}, though there are many extensions as well. For adversarial cyber-physical systems there is differential game logic in~\cite{DBLP:journals/tocl/Platzer15,DBLP:journals/tocl/Platzer17}, and Part 5 of~\cite{Platzer18}. There are also extensions for distributed hybrid systems (quantified differential dynamic logic,~\cite{DBLP:conf/csl/Platzer10}), stochastic hybrid systems (stochastic differential dynamic logic,~\cite{DBLP:conf/cade/Platzer11}), differential algebraic programs (differential-algebraic dynamic logic,~\cite{DBLP:journals/logcom/Platzer10}), and a temporal extension of \DDL{} called differential temporal dynamic logic~\cite{DBLP:conf/lfcs/Platzer07},~\cite[Chapter 4]{Platzer10}.

In addition to verification of hybrid systems, the present work falls more generally into the category of formal verification or simulation of logical systems inside theorem provers. PVS0 is an embedding of a fragment of the specification language of PVS \textit{within} PVS, used in termination analysis of recursive functions~\cite{MAMDNAAF21}. Other efforts to model or verify theorem provers  include work on the prover kernel of Hol Light~\cite{harrison2006towards}, the type-checker of Coq~\cite{sozeau2019coq}, the soundness of ACL2~\cite{davis2015reflective}.
The goal of \name{} is to add to hybrid systems reasoning to the toolbox of PVS increasing its proving capabilities. PVS has been used in verification projects in domains such as aircraft avoidance systems~\cite{munoz2015daidalus}, path planning algorithms~\cite{colbert2020polysafe}, unmanned aircraft systems~\cite{munoz2016unmanned}, position reporting algorithms of aircraft~\cite{dutle2021formal}, sensor uncertainty mitigation~\cite{narkawicz2018sensor},  floating point error  analysis~\cite{moscato2017automatic,TitoloFMM18}, genetic algorithms~\cite{nawaz2013formal}, nonlinear control systems~\cite{bernardeschi2016verifying}, and requirements written in linear temporal logic and FRETish~\cite{ConradTGPD2022}. In addition to advanced real number reasoning capabilities~\cite{daumas2008verified,munoz2013formalization,NM14,moscato2015affine,narkawicz2013formally,narkawicz2015formally} provided by PVS, previous work has connected PVS to the automated theorem prover MetiTarski~\cite{akbarpour2010metitarski}, for automated reasoning of universally quantified statements about real numbers, including several transcendental functions~\cite{denman2014automated}.  The capability to use MetiTarski in PVS is leveraged  in the \dlref{dl-assert} command in \name.  \section{Conclusion and future work} \label{sec:con}
This work describes \name, a logical embedding of \DDL{} in PVS. This embedding extends the formal verification abilities of PVS by giving a framework for specifying and reasoning about HPs and allowing features of PVS to be used naturally within the \DDL{} embedding. Novel features include support for importing user-defined functions and theories such as the extensive math and computer science developments available in NASAlib. Additionally, this embedding allows for meta reasoning about HPs and \DDL{} at the PVS level. An example was given that illustrates capabilities of Plaidypvs that go beyond what could be
be accomplished in a stand-alone implementation of \DDL{} alone such as KeYmaera~X.

Regarding future work, one natural next step is to apply \name{} to safety-critical applications of interest to NASA. This will include formal verification of hybrid systems related to urban air mobility and wildland fire fighting among others. Another direction is to increase the usability of \name. To do so, a Visual Studio Code extension is under development to display specifications and the proof calculus in a natural and user-friendly way. To increase the automation of \DDL{} within \name{}, a more complete substitution calculus to include Boolean expressions containing statements about hybrid programs will be implemented. Additionally, formal verification of liveness properties is intended, with implementations of strategies to match. Furthermore, a more robust ordinary differential equation solver to enhance the capabilities of the \emph{differential solve} command would increase the usability of \name{} greatly. Finally, a detailed description of the multivariate analysis and ordinary differential equation library developed to support this embedding will be written similar to the semi-algebraic set library (\cite{slagel2021formal}), which was done to support verification of liveness properties in upcoming work. 

The semantic structure of \DDL{} in \name{} is based on the input/output semantics. Future work on defining the trace semantics of hybrid programs will extend the analysis capabilities of the embedding, such as being able to define properties in linear temporal logic like the work in~\cite{jeannin2014dtl}. It has been noted that quantifier elimination, is often the bottleneck for formal verification of hybrid programs, due to the computational complexity of the general problem. Implementation of techniques to make this process faster would help the usability of \name{}. There are many directions to go for this effort, but one direction will be implementation of the active corners method for a specific class of quantifier elimination~\cite{kheterpal2022automating} geared towards formalized reasoning of aircraft operations.

\bibliographystyle{eptcs}

\end{document}